\documentclass[letterpaper,iop,apj]{emulateapj}
\pdfoutput=1
\usepackage{thumbpdf}
\usepackage{amsmath,amssymb}
\usepackage{graphicx,mathptmx}
\usepackage{natbib}
\usepackage[usenames]{color}
\usepackage{ifpdf}
 \usepackage{xcolor}[2007/01/21]
\usepackage{refcount}
\usepackage{textcase}
\usepackage{epstopdf}

\ifpdf
\pdfoutput=1 
\else

\fi

\newcommand{\beq}{
\begin{equation}
}
\newcommand{\eeq}{
\end{equation}
}
\newcommand{\beqa}{
\begin{eqnarray}
}
\newcommand{\eeqa}{
\end{eqnarray}
}
\newcommand{\units}[1]  {\ensuremath{\mathrm{\ {#1}}}}
\newcommand{\msun}     {\ensuremath{{{M}}_{\scriptscriptstyle \odot}}}
\newcommand{\lsun}     {\ensuremath{{{L}}_{\scriptscriptstyle \odot}}}
\newcommand{\lsunv}     {\ensuremath{{L}}_{{\scriptscriptstyle \odot},V}}
\newcommand{\kms}      {\ensuremath{~\mathrm{km~s^{-1}}}}

\newcommand{\ergs}     {\ensuremath{~\mathrm{erg~s^{-1}}}}

\newcommand{\msigma}   {\ensuremath{M}{--}\ensuremath{\sigma}}
\newcommand{\ml}       {\ensuremath{M}{--}\ensuremath{L}}
\newcommand{\mbh}      {\ensuremath{M_{\mathrm{BH}}}}

\providecommand{\ion}[2]{#1$\;$\textsmaller{\@Roman{#2}}}

\newcommand{\figwidth}{0.475\textwidth}
\newcommand{\figwidthtwo}{0.35\textwidth}

\def\spose#1{\hbox to 0pt{#1\hss}}
\newcommand{\lta}{\mathrel{\spose{\lower 3pt\hbox{$\mathchar"218$}}
      \raise 2.0pt\hbox{$\mathchar"13C$}}}
\newcommand{\gta}{\mathrel{\spose{\lower 3pt\hbox{$\mathchar"218$}}
      \raise 2.0pt\hbox{$\mathchar"13E$}}}
\def\simlt{\mathrel{\rlap{\lower 3pt\hbox{$\sim$}}\raise 2.0pt\hbox{$<$}}}
\def\simgt{\mathrel{\rlap{\lower 3pt\hbox{$\sim$}} \raise 2.0pt\hbox{$>$}}}

\definecolor{KayhanCiteColor}{rgb}{0,0.08,0.35}
\definecolor{KayhanURLColor}{rgb}{0,0.08,0.35}
\definecolor{KayhanLinkColor}{rgb}{0,0.08,0.35}
\definecolor{KayhanPageColor}{rgb}{0,0.08,0.35}
\definecolor{medred}{rgb}{0.75,0.0,0.0}

\shorttitle{BH in NGC 4382?}
\shortauthors{G\"{u}ltekin et al.}

\submitted{Accepted by The Astrophysical Journal}

\makeatletter
\ifx\undefined\hyperref
\usepackage[pdftitle={Is There a Black Hole in NGC 4382?},pdfauthor={Kayhan Gultekin},pdfsubject={Astrophysics},pdfkeywords={galaxies: individual (NGC 4382, M85) --- galaxies: kinematics and dynamics --- black hole physics --- galaxies: nuclei},letterpaper,linktocpage,colorlinks=true,linkcolor={KayhanLinkColor},urlcolor={KayhanURLColor},citecolor={KayhanCiteColor},hyperfootnotes=false]{hyperref}
\else\relax\fi
\makeatother
\usepackage[figure,figure*]{hypcap}
\usepackage{atveryend}
\usepackage[
  open,
  openlevel=3,
  atend
]{bookmark}[2010/04/08]
\begin{document}

\label{firstpage}
 
\title{Is There a Black Hole in NGC 4382?\footnotemark[1]}
\footnotetext[1]{Based on observations made with the \emph{Hubble
Space Telescope}, obtained at the Space Telescope Science Institute,
which is operated by the Association of Universities for Research in
Astronomy, Inc., under NASA contract NAS 5-26555.  These observations
are associated with GO proposals 5999, 6587, 6633, 7468, and 9107.}

\author{Kayhan G\"{u}ltekin}
\author{Douglas O.\ Richstone}
\affil{University of Michigan, Ann Arbor, MI, 48109;
\href{mailto:kayhan@umich.edu}{kayhan@umich.edu}.}
\author{Karl Gebhardt}
\affil{Department of Astronomy, University of Texas, Austin, TX, 78712.}
\author{S.~M.\ Faber}
\affil{University of California Observatories/Lick Observatory, Board
of Studies in Astronomy and Astrophysics, University of California,
Santa Cruz, CA 95064.}
\author{Tod R.\ Lauer}
\affil{National Optical Astronomy Observatory, Tucson, AZ 85726.}
\author{Ralf Bender}
\affil{Universit\"ats-Sternwarte M\"unchen,
Ludwig-Maximilians-Universit\"at, Scheinerstr. 1, D-81679.}
\author{John Kormendy}
\affil{Department of Astronomy, University of Texas, Austin, TX, 78712.}
\author{Jason Pinkney}
\affil{Department of Physics and Astronomy, Ohio Northern University,
Ada, OH 45810.}

\begin{abstract}
\hypertarget{abstract}{}%
We present \emph{HST} STIS observations of the galaxy NGC 4382 (M85)
and axisymmetric models of the galaxy to determine mass-to-light ratio
($\Upsilon_V$) and central black hole mass (\mbh).  We find
$\Upsilon_V = 3.74\pm 0.1\ \msun/\lsun$ and $\mbh =
1.3^{+5.2}_{-1.2}\times10^7\ \msun$ at an assumed distance of 17.9
Mpc, consistent with no black hole.  The upper limit, $\mbh < 9.6
\times 10^7\ \msun (2\sigma)$ or $\mbh < 1.4 \times 10^8 (3\sigma)$ is
consistent with the current \msigma\ relation, which predicts $\mbh =
8.8 \times 10^7\ \msun$ at $\sigma_e = 182\kms$, but low for the
current \ml\ relation, which predicts $\mbh =7.8 \times 10^8\ \msun$
at $L_V = 8.9 \times 10^{10}\ \lsunv$.  \emph{HST} images show the
nucleus to be double, suggesting the presence of a nuclear eccentric
stellar disk, in analogy to the Tremaine disk in M31.  This conclusion
is supported by the \emph{HST} velocity dispersion profile.  Despite
the presence of this non-axisymmetric feature and evidence of a recent
merger, we conclude that the reliability of our black hole mass
determination is not hindered.  The inferred low black hole mass may
explain the lack of nuclear activity.  
\bookmark[ rellevel=1,
keeplevel, dest=abstract ]{Abstract}
\end{abstract}

\section{Introduction}
\label{intro}

Finding a black hole at the center of a galaxy is no longer a
surprise.  The prevalence of these black holes is well established
\citep{richstoneetal98}.  Their importance has also been recognized,
for example, as active galactic nuclei central engines \citep{rees84}.
The tight correlation of black hole masses with host galaxy properties
strongly suggests an underlying link between galaxy and black hole
evolution.  The black hole mass has been found to be correlated with
the stellar spheroid's mass \citep{dressler89, magorrianetal98,
2001ApJ...553..677L, 2002MNRAS.331..795M, mh03, 2004ApJ...604L..89H},
luminosity \citep[the \ml\ relation,][]{kormendy93a, kr95,
magorrianetal98, kg01, 2009ApJ...698..198G}, stellar velocity
dispersion \citep[the \msigma\ relation,][]{fm00, gebhardtetal00a,
tremaineetal02, 2009ApJ...698..198G}, galaxy core parameters
\citep{milosavljevicetal02, 2002ApJ...566..801R, 2004ApJ...613L..33G,
ferrareseetal06, 2006ApJ...648..976M, laueretal07b,
2009ApJ...691L.142K}, and globular cluster system
\citep{2010ApJ...720..516B}.  Current theoretical work addressing
these scaling relations focuses on feedback from outflows that are
powered by accretion onto the central black hole.  Thus the
consequences of black hole growth plays a role in regulating star
formation in the galaxy \citep[e.g.,][]{2006MNRAS.370..645B,
2006ApJS..163....1H}.  Despite this progress, we do not have a
complete understanding of the physics involved, and the pursuit of
black hole masses remains important.

Masses found from primary, direct, dynamical measurements are the
basis from which all other black hole masses are derived.  Indirect
mass indicators, such as AGN line widths, are calibrated to
reverberation mapping, direct yet secondary measurements
\citep{bentzetal06}, which are themselves normalized against the
direct dynamical measurements \citep{onkenetal04,
2010ApJ...716..269W}.  Currently, even the empirical scaling relations
are incomplete.  The \ml\ and \msigma\ relations make different
predictions at the upper end \citep{laueretal07}.  The possibility of
increased intrinsic scatter at the low end remains untested
\citep{volonteri07}, and there is growing evidence that late-type
galaxies as a whole and pseudo-bulges in particular do not lie on the
same relation as early type galaxies \citep{2008MNRAS.386.2242H,
2009ApJ...698..198G, 2010arXiv1007.2851G, kbcnat10}.  There is also
much interest in multi-parameter extensions to these relations
\citep{2007ApJ...665..120A, 2007ApJ...669...67H}.

\label{selection}
It is this last open question, the existence and utility of
multi-parameter scaling relations, for which the galaxy in this study,
\object[M85]{NGC 4382} (M85), was selected.  NGC 4382 lies in a narrow
range in velocity dispersion ($180 < \sigma < 220\kms$, for which
$\mbh \sim 10^8\ \msun$) based on HyperLEDA\footnote{Available at
\href{http://leda.univ-lyon1.fr/}{http://leda.univ-lyon1.fr/}.}
central velocity dispersion measures \citep{hyperleda}.  With enough
galaxies from a narrow range in velocity dispersion, we may test for
additional trends in black hole mass with other host galaxy
parameters.  This range was chosen because galaxies in this range may
have either core or power-law surface brightness profiles and because
both late-type and early-type galaxies lie in this range.  The
galaxies were also selected based on their distances so that the
predicted radius of influence was larger than 0\farcs1.  The radius of
influence is defined as
\beq
   R_{\mathrm{infl}} \equiv \frac{G \mbh}{\sigma^2\left(R_{\mathrm{infl}}\right)},
\label{e:rinfdef}
\eeq
where the velocity dispersion $\sigma(R)$ is a function of projected
distance from the center along the major axis and is evaluated at the
radius of influence.  Black hole masses were estimated from their
central velocity dispersion measurement and the \msigma\ fit due to
\citet{tremaineetal02}.

NGC 4382 is an E2 galaxy \citep{2009ApJS..182..216K} with diffuse
stellar light surrounding it, which has led some to classify it as an
S0 \citep{rc3}.  We take the distance to NGC 4382 to be 17.9\ Mpc
(calculated assuming a Hubble constant of $H_0 =
70~\kms~\mathrm{Mpc^{-1}}$).  The surface brightness profile as a
function of radius reveals a core at the center
\citep{2005AJ....129.2138L} and may be parameterized with a ``Nuker
Law'' given by
\beq I\left(r\right) = 2^{\left(\beta - \gamma\right)/\alpha} I_b 
   \left(\frac{r_b}{r}\right)^\gamma \left[1 + \left(\frac{r}{r_b}\right)^\alpha\right]^{\left(\gamma - \beta\right)/\alpha},
\label{e:nukerlaw}
\eeq
which is a broken power-law profile with variable sharpness in the
break \citep{1995AJ....110.2622L}.  NGC 4382 has $r_b = 0\farcs93 =
80.7\units{pc}$, $I_b = 15.67\units{mag\ arcsec^{-2}}$, $\alpha =
1.13$, $\beta = 1.39$, and $\gamma = 0.00$
\citep{2005AJ....129.2138L}.  The total luminosity of the galaxy is
$M_V = -22.54$.


In section \ref{obs} we describe the observations and data reduction,
including new space spectroscopic observations (section \ref{stis}),
ground-based spectra (section \ref{grspectra}), and imaging data
(section \ref{imaging}).  The kinematic modeling and its results are
presented in section \ref{model}, and we discuss the caveats for and
implications of our results in section \ref{concl}.

\section{Observations} 
\label{obs}

\subsection{STIS Observations and Data Reduction}
\label{stis}
Measuring black hole masses precisely requires spectra with high
spatial resolution.  Thus, most precise black hole mass measurements
come from observations using the \emph{Hubble Space Telescope}
(\emph{HST}), though adaptive optics techniques are showing promise
\citep[e.g.,][Gebhardt et al.\ in preparation]{2009MNRAS.399.1839K,
2010MNRAS.403..646N} and mass measurements using maser observations are
ramping up \citep{2010arXiv1008.2146K}.  We observed \ion{Ca}{2}
triplet absorption from NGC 4382 with the Space Telescope Imaging
Spectrograph (STIS) on \emph{HST} set with the G750M grating and a
$52\arcsec\times0\farcs2$ slit.  The medium-dispersion grating, as
opposed to the low-dispersion G750L grating, is necessary in order to
get the spectral resolution high enough to recover line-of-sight
velocity distributions (LOSVDs) with sufficient precision.  The slit
was set at a width of 0\farcs2 to optimize the signal-to-noise ratio
as core ellipticals have relatively low central surface brightness and
widening the slit as far as possible allows as short an observation as
possible.  The slit was positioned along at a position angle of
$\mathrm{PA} = 48^\circ$ east of north close to the the photometric
major axis position angle of $\mathrm{PA} = 30^\circ$ as determined
from \emph{HST}/WFPC2 observations \citep[see section \ref{spaceimage}
below,][]{2005AJ....129.2138L, 2009ApJS..182..216K}.  
We obtained 16 exposures at 5 dither positions for a total of exposure
of 18\,794\units{s}.  The STIS CCD has a 1024$\times$1024 pixel
format, a readout noise of $\sim 1 e^-\ \mathrm{pix}^{-1}$, and a gain
of 1.0 without on-chip binning.  The spectra spanned a wavelength
range of 8257--8847\ \AA, and our wavelength solutions revealed a
reciprocal dispersion of 0.554\ \AA\ pix$^{-1}$, and the spatial scale
was 0\farcs05071 pix$^{-1}$ for G750M at 8561 \AA.

The STIS data reduction was done with routines developed for this
purpose \citep{pence98,pinkneyetal03} following the standard pipeline:
Raw spectra were extracted from the multi-dimensional FITS file, and
then a constant fit to the overscan region was subtracted to remove
the bias level.  The STIS CCD has warm and hot pixels that change on
timescales of about a day.  These pixels require that the subtraction
of dark current be accurate, which we accomplished by using the
iterative \emph{self-dark} technique \citep{pinkneyetal03}.  After
flat-fielding and dark subtraction, spectra were shifted vertically to
a common dither, combined, and rotated.  One-dimensional spectra were
then extracted using a bi-weight combination of rows.  Near the galaxy
center, we adopted a 1-pixel wide binning for maximum spatial
resolution.  These data-reduction methods are similar to those in
\citet{pinkneyetal03}, which the interested reader may consult for
details.

LOSVDs were extracted from reduced spectra as described in
\citet{2009ApJ...695.1577G}.  Each galaxy spectrum is a convolution of
the intrinsic spectrum of stars observed in the aperture with the
LOSVD of those stars.  We deconvolved the observed galaxy spectrum
using the template spectrum composed from standard stellar spectra
using a maximum penalized-likelihood method
\citep{gebhardtetal03,pinkneyetal03}.  We present Gauss-Hermite
moments of the extracted velocity profiles in Table
\ref{t:n4382stisdata} and Figure \ref{f:n4382spec}.  Although it is
common practice to communicate velocity profiles with Gauss-Hermite
moments, we use LOSVDs binned in velocity space for our modeling
described below in section \ref{model}.

\begin{deluxetable}{rrr@{$\pm$}lr@{$\pm$}lr@{$\pm$}lr@{$\pm$}l}
  \footnotesize
  \tablecaption{Kinematic Profile for NGC 4382 from STIS Observations}
  \tablehead{
     \colhead{$R$} &
     \colhead{Width} &
     \multicolumn{2}{c}{$V$} &
     \multicolumn{2}{c}{$\sigma$} &
     \multicolumn{2}{c}{$h_3$} &
     \multicolumn{2}{c}{$h_4$} \\
     \colhead{(\arcsec)} &
     \colhead{(pix)} &
     \multicolumn{2}{c}{($\mathrm{km\ s^{-1}}$)} &
     \multicolumn{2}{c}{($\mathrm{km\ s^{-1}}$)} &
     \multicolumn{2}{c}{} &
     \multicolumn{2}{c}{} 
  }
  \startdata
0.00 & 1 &  15 & 17 & 147 & 20 & $-$0.017 & 0.06 & $-$0.068 & 0.060\\
0.05 & 1 &   4 & 14 & 162 & 12 & $-$0.049 & 0.05 & $-$0.045 & 0.036\\
0.10 & 1 &  24 & 15 & 150 & 13 & $-$0.006 & 0.05 & $-$0.067 & 0.029\\
0.18 & 2 &  31 & 15 & 145 & 12 & $-$0.071 & 0.05 & $-$0.049 & 0.032\\
0.30 & 3 &  14 & 16 & 170 & 17 &    0.021 & 0.06 & $-$0.063 & 0.046\\
0.58 & 8 &  24 & 18 & 178 & 20 & $-$0.021 & 0.07 & $-$0.016 & 0.053\\
1.12 &14 &  20 & 19 & 166 & 20 & $-$0.001 & 0.07 &    0.015 & 0.060
\enddata
\label{t:n4382stisdata}
\tablecomments{Gauss--Hermite moments for velocity profiles derived
from STIS data.  First and second moments are given in units of $
\mathrm{km\;s^{-1}}$.  Radii are given in arcsec, and the second
column gives the width of the radial bin in pixels, which are
0\farcs051.}
\end{deluxetable}
\bookmarksetup{color=[rgb]{0,0,0.54}}
\bookmark[
rellevel=1,
keeplevel,
dest=table.\getrefnumber{t:n4382stisdata}
]{Table \ref*{t:n4382stisdata}: Gauss-Hermite moments of NGC 4382}
\bookmarksetup{color=[rgb]{0,0,0}}

\subsection{Ground-Based Spectra}
\label{grspectra}

Ground-based velocity information was obtained from archival data
using Spectrographic Areal Unit for Research on Optical Nebulae
(SAURON)%
\footnote{Downloaded from
\href{http://www.strw.leidenuniv.nl/sauron/}{http://www.strw.leidenuniv.nl/sauron/}
in Feb 2010.}, an integral-field spectrograph unit mounted on the
William Herschel Telescope in La Palma \citep{2001MNRAS.326...23B}.
As the SAURON instrument and observations were designed to measure and
characterize the internal kinematics of the selected galaxies
\citep{2002MNRAS.329..513D}, the data are excellent for our purposes.
In its low resolution mode, the instrument has a field of view of
$33\times41\arcsec$ with 0\farcs94 pixels, each of which provides a
spectrum with $\mathrm{FWHM} = 4.2\ $\AA\ spectral resolution.  This
spectral resolution corresponds to $\sigma_\mathrm{instr} = 108\kms$
at $4950\ $\AA, near the center of the 4800--5380\ \AA\ wavelength
range.  The data were taken on 14 March 2001 in two pointings, each
consisting of 4 exposures of 1800\ s each under FWHM$ = 2\farcs7$
seeing \citep{emsellemetal04}.

The stellar kinematics data are provided as Gauss-Hermite moments for
each lenslet position on the galaxy.  The Gauss-Hermite moments from
each lenslet were converted into LOSVDs.  The data from each quadrant
of the galaxy were combined into one, changing signs of the odd
moments as necessary.  We binned the data into 4 position angles (0,
20, 30, and 70$^{\circ}$ east of the major axis) with 7 radial bins
and one position angle (45$^{\circ}$ east of the major axis) with 6
radial bins for a total of 34 LOSVDs from ground-based data.  

\begin{figure*}[tp]
\hypertarget{figspec}{}%
\centering
\includegraphics[width=0.7\textwidth]{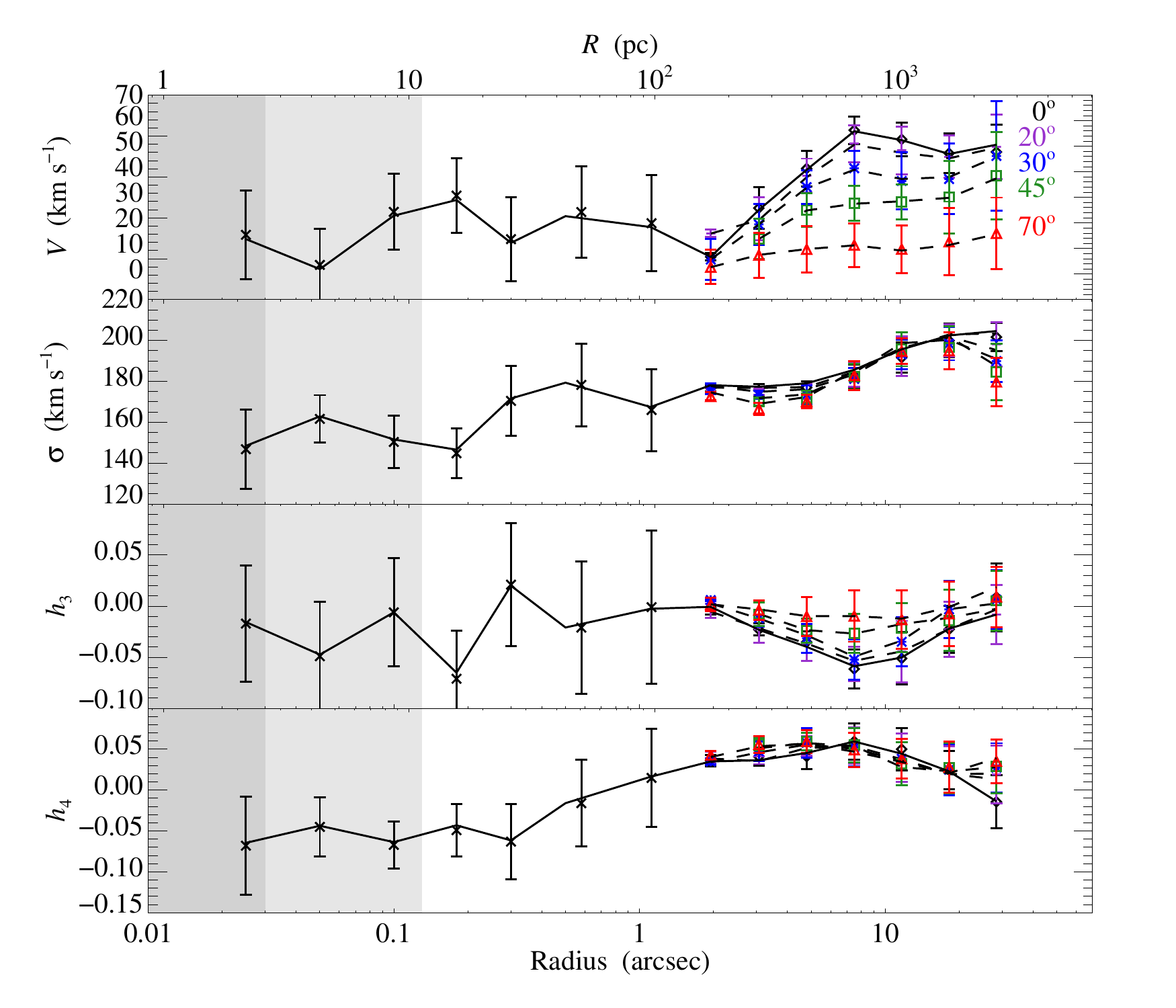}
\caption{Gauss--Hermite moments of LOSVDs for NGC 4382 as a function
of radius, from top to bottom: $V$, $\sigma$, $h_3$, and $h_4$. Values
at the center of the galaxy ($R=0$) are plotted at $R = 0\farcs025$.
Moments from opposite sides of the galaxy are symmetrized and
combined.  The sense of the sign is such that radius is increasing to
the southwest.  Crosses are Gauss-Hermite moments of LOSVDs from
\emph{HST} STIS data along the major axis.  Also plotted are
ground-based Gauss-Hermite moments of LOSVDs along the major axis
(black diamonds) and skew axes (other colors with PA relative to the
major axis indicated on the right in the top panel) from SAURON
archives.  Though Gauss-Hermite moments are not fit directly in the
modeling, the jagged lines are the resulting Gauss-Hermite fit to the
best-fit model's LOSVDs from section \protect{\ref{model}} for the
major axis (solid) and skew axes (dashed).  The best-fit model has
$M_{\mathrm{BH}} = 1.5\times10^7~\msun$ and $\Upsilon_V = 3.72$.  We
indicate the radius of the sphere of influence of a black hole with
mass predicted by the \msigma\ relation (light gray) or with our
best-fit mass (medium gray).}
\label{f:n4382spec}
\end{figure*}

To incorporate errors in the Gauss-Hermite moments in our data set we
created $10^4$ Monte Carlo realizations of LOSVDs for each SAURON
lenslet.  As described in \citet{2009ApJ...695.1577G}, Gauss-Hermite
moments corresponding to unphysical negative values are assigned a
value of zero with a conservative uncertainty.  For each spatial bin,
we took the median LOSVD of all realizations at a given velocity and
the standard deviation as the error, which dominated the individual
measurement errors.  The Gauss-Hermite moments of the LOSVDs are
presented in Figure~\ref{f:n4382spec} as a function of radius along
with our STIS data.  These measurements agree well with the
ground-based kinematics from \citet{fisher97}.

The profiles were binned into 13 equal bins in velocity from
$-500$--$500$\kms\ about the systemic velocity covering the range of
velocities measured.  Using the ground kinematic data, we compute
an effective stellar velocity dispersion, defined as
\beq
\sigma^2_e \equiv \frac{\int_{0}^{R_e} \left({\sigma(r)^2 + V(r)^2}\right) I\left(r\right) dr}{{\int_{0}^{R_e} I\left(r\right) dr}},
\label{e:sigmae}
\eeq
where $R_e$ is the effective radius, $I(r)$ is the surface brightness
profile (see section \ref{imaging} below), and $V(r)$ and $\sigma(r)$
are the first and second Gauss-Hermite moments of the LOSVD.  Using a
non-parametric method of integrating the brightness and ellipticity
profile \citet{2009ApJS..182..216K} find $R_e = 102 \pm 6\arcsec$.
From the ground-based velocity profile, we find an effective stellar
velocity dispersion of $\sigma_e = 182 \pm 5\kms$.

\bookmarksetup{color=[rgb]{0.54,0,0}}
\bookmark[rellevel=1,keeplevel,dest=figspec]{Fig \ref*{f:n4382spec}: Observed and best-fit LOSVDs}
\bookmarksetup{color=[rgb]{0,0,0}}

\subsection{Imaging}
\label{imaging}
\label{spaceimage}
\label{grimage}

\begin{figure}
\hypertarget{figsurfb}{}%
\centering
\includegraphics[width=\figwidth]{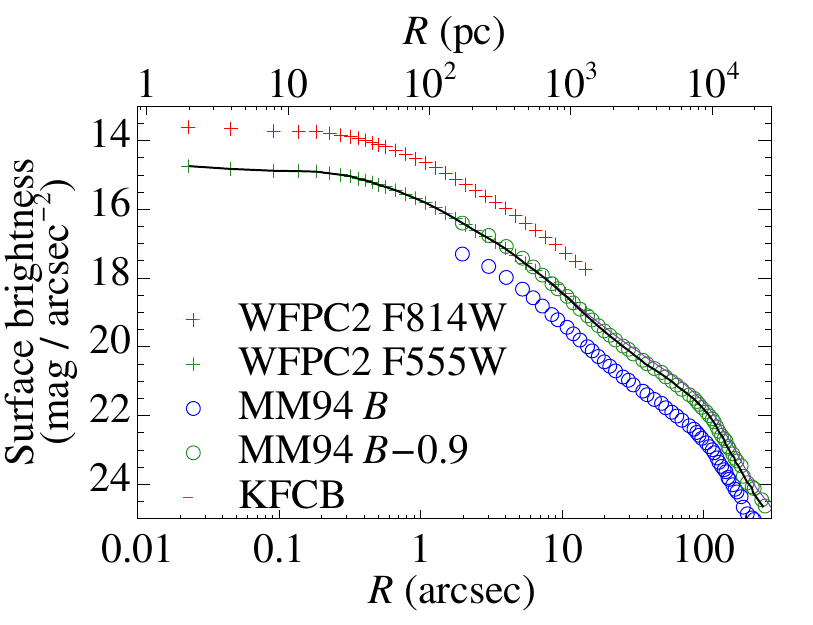}
\caption{Surface brightness of NGC 4382.  Red and green crosses (upper and
lower) show F814W and F555W WFPC2 data, respectively, from
\citet{2005AJ....129.2138L}.  Blue circles show
\citet{1994A&AS..105..481M} ground photometry from the 1.2-m telescope
of the Observatoire de Haute-Provence with a PSF FWHM of 3\farcs12.
Green circles are the same data but color-shifted by 0.9 magnitudes to
match the F555W data.  The solid curve is our assumed $V$-band surface
brightness profile of NGC 4382.  The purple line is data from
\citet{2009ApJS..182..216K} and comes from a variety of instruments
and bands.  The difference between our adopted surface brightness
profile and the purple line is small and can be attributed primarily
to slightly different colors at large radii.}
\label{f:n4382surfb}
\end{figure}

The high-resolution photometry of the central regions of NGC 4382 comes
from Wide Field Planetary Camera 2 (WFPC2) observations on \emph{HST}
using filters F555W ($V$) and F814W ($I$).  The observations, data
reduction, and surface brightness profiles (including Nuker profile
fits) are detailed by \citet{2005AJ....129.2138L}. Surface brightness profiles
are also available at the Nuker web page.\footnote{See
\href{http://www.noao.edu/noao/staff/lauer/wfpc2\_profs/}{http://www.noao.edu/noao/staff/lauer/wfpc2\_profs/}.}
The wide-field data that we use are literature data from ground
observations on the 1.2-m telescope of the Observatoire de
Haute-Provence with PSF FWHM$ = 3$\farcs12 originally obtained by
\citet{1994A&AS..105..481M}.  The ground data are $B$-band data,
which we have color-corrected with $B-V = 0.9$, determined by
requiring the space data and ground data to match where they overlap
in the spatial direction.  Figure \ref{f:n4382surfb} shows the surface
brightness profiles of the space and ground data along with our
adopted, combined surface brightness profile for our modeling.

Since our modeling efforts began, a superior ground-based photometric
data set became available \citep{2009ApJS..182..216K}.  The data come
from several different instruments.  Inside of 1\arcsec, it is the
same WFPC2 F555W data that we use.  Starting at 1\arcsec, the profile
is an average of WFPC2 and ACS data as well as two different cameras
on the Canada-France-Hawaii Telescope.  At the largest radii, the data
come from the McDonald 0.8-m telescope, which has a wide field of view.
We plot it in Fig. \ref{f:n4382surfb} for comparison.  The agreement
inside of $R < 15\arcsec$ is striking.  At large radii, there is a
small but systematic deviation from our adopted surface brightness
profile.  This is almost certainly due to the different colors used
and is likely evidence of a slight color gradient starting outside of
$R > 100\arcsec$.  At these large radii, the smaller PSF of the
\citet{2009ApJS..182..216K} data set does not have a large advantage,
and the extremely wide coverage---out to $R=625\arcsec$---is not used
since our kinematic profiles only go out to $R=30\arcsec$.

\bookmarksetup{color=[rgb]{0.54,0,0}}
\bookmark[rellevel=1,keeplevel,dest=figsurfb]{Fig \ref*{f:n4382surfb}: Surface-brightness profile}
\bookmarksetup{color=black}

\section{Kinematic Modeling}
\label{model}
\label{n4382}

  We use the three-integral, axisymmetric Schwarzschild method to make
kinematic models of NGC 4382.  The model is constructed in several steps.
First the observed surface-brightness profile, $\Sigma(r)$, is
deprojected into an axisymmetric luminosity density, $j(r,\theta)$.
This deprojection assumes that surfaces of constant luminosity density
are coaxial ellipsoids and depends on a chosen inclination, which we
take to be $i = 90^\circ$.  Under the assumption of constant
mass-to-light ratio of a chosen value, $\Upsilon$, the mass density of
stars is trivially obtained $\rho(r,\theta) = \Upsilon j$.  We can
then calculate the stellar gravitational potential from Poisson's
equation.  The potential from a point mass with a chosen mass, $\mbh$,
is then added at $r = 0$.

With the potential for the entire system in hand, we then calculate
orbits of representative stars.  The number of orbits is different for
each mass model with more orbits needed to sample the phase space when
$\Upsilon$ or $\mbh$ is large, but the number ranged from 15735 to
17213 orbits.  The number of orbits is increased by increasing the
density of orbits in energy, angular momentum, and non-classical third
integral space.  We increase the number of orbits with sparsely
gridded parameter space (\mbh and $\Upsilon_V$) until our results
converge, and then we run a finer grid in parameter space, which we
report here.  The amount of time each representative star spends in a
given bin in position and velocity space is monitored so as to
identify the orbits' possible contribution to the observed surface
brightness and LOSVD.  For each set of parameters, we find the
non-negative weights and a goodness-of-fit ($\chi^2$) for the set of
orbits that best matches the observed LOSVD while reproducing the
observed surface brightness.  The entire method is explained in more
detail in \citet{gebhardtetal03} and \citet{siopisetal08}.

\begin{figure}
\hypertarget{figchisq}{}%
\centering
\includegraphics[width=\figwidthtwo,angle=90]{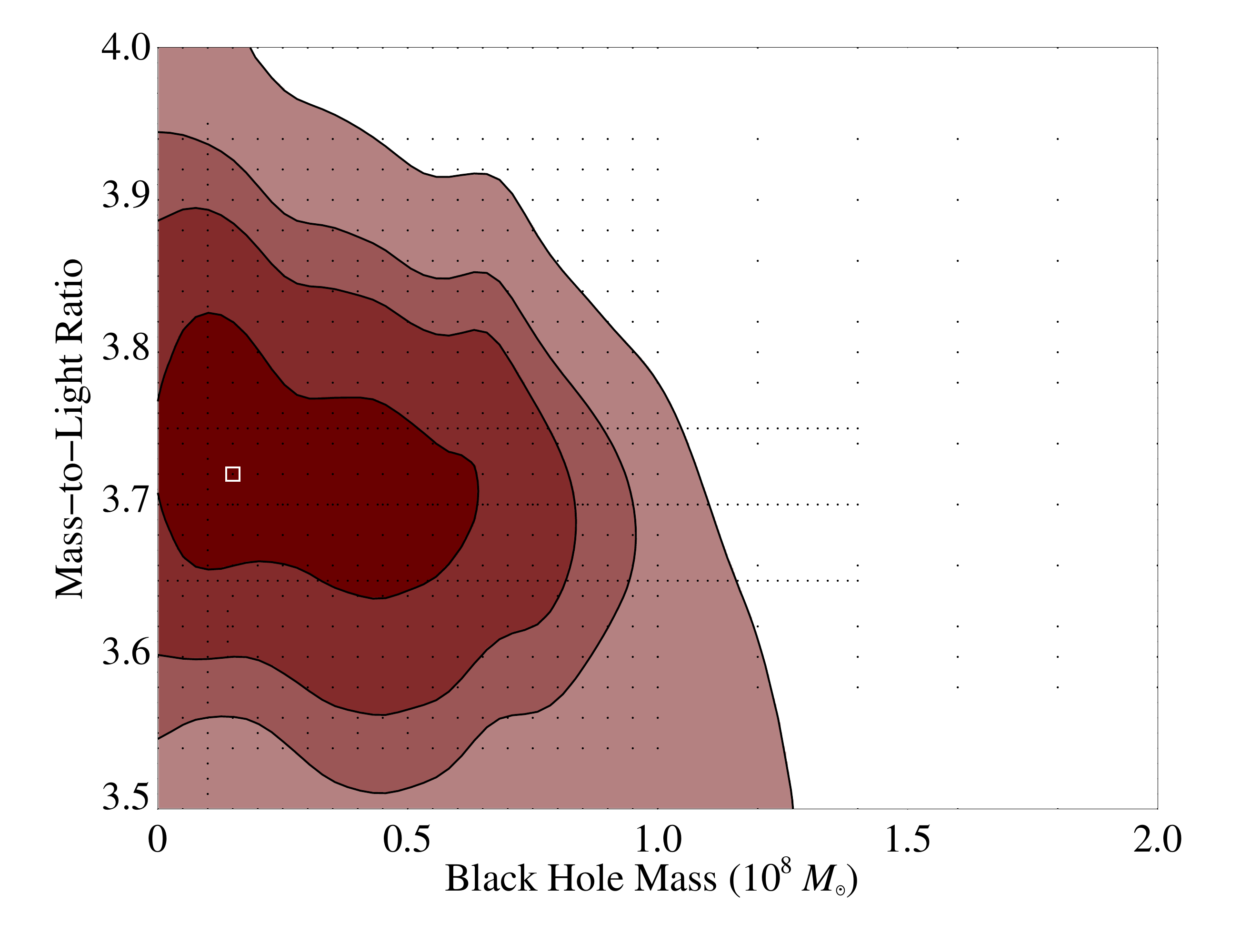}
\caption{Mass modeling $\chi^2$ contours for NGC 4382.  Contours are
  for $\Delta\chi^2 =$ 1.00, 2.71, 4.00, and 6.63, which bracket
  individual parameter confidence levels of 68.3, 90.0, 95.4, and
  99.0\%, respectively.  Contours have been smoothed for plotting, and
  each level is filled with a solid color.  The square shows the
  best-fit model.  Dots indicate parameters modeled.  The best-fit
  model has $M_{\mathrm{BH}} = 1.5\times10^7~\msun$ and $\Upsilon_V =
  3.72\ \msun / \lsunv$.  Marginalizing over the other parameter we
  find $M_{\mathrm{BH}} = 1.3^{+5.2}_{-1.2}\times10^7~\msun$ and
  $\Upsilon_V = 3.74\pm 0.10\ \msun / \lsunv$.}
\label{f:n4382chi}
\end{figure}

\begin{figure}
\hypertarget{figmargm}{}%
\centering
\includegraphics[width=\figwidthtwo]{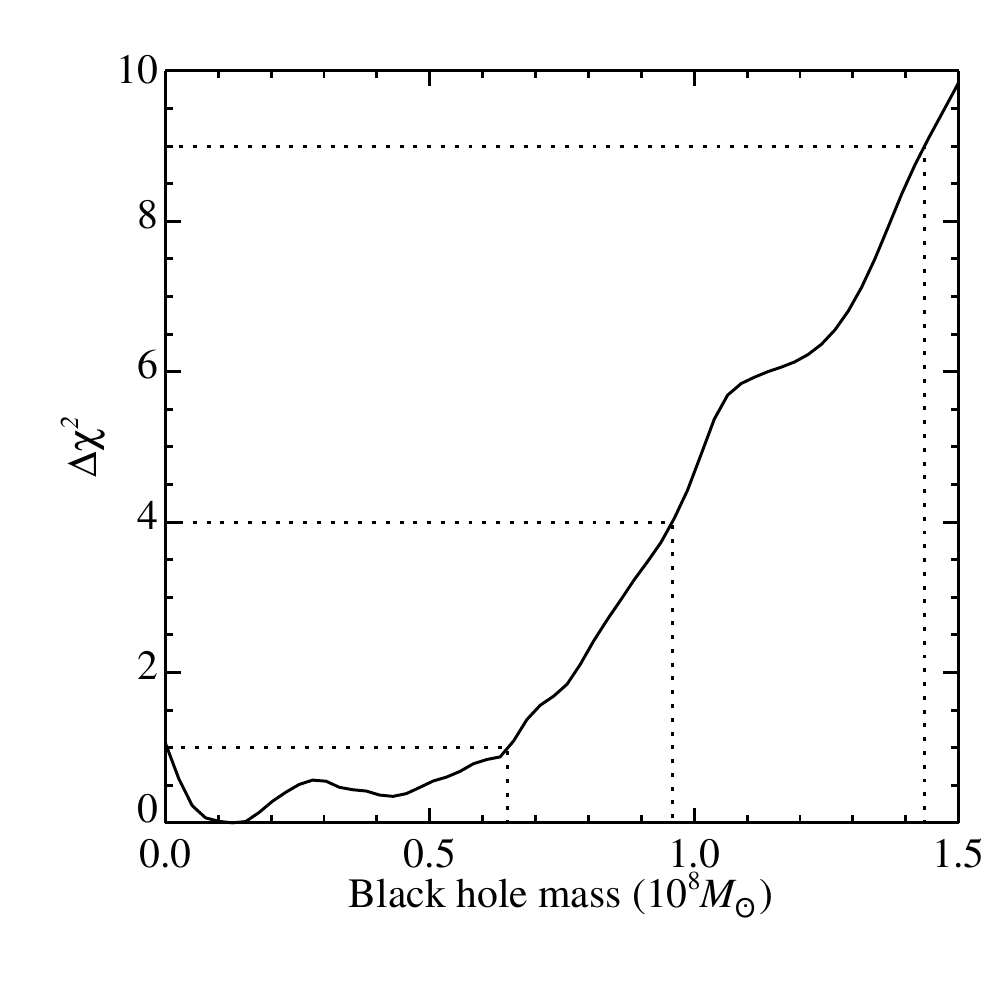}
\includegraphics[width=\figwidthtwo]{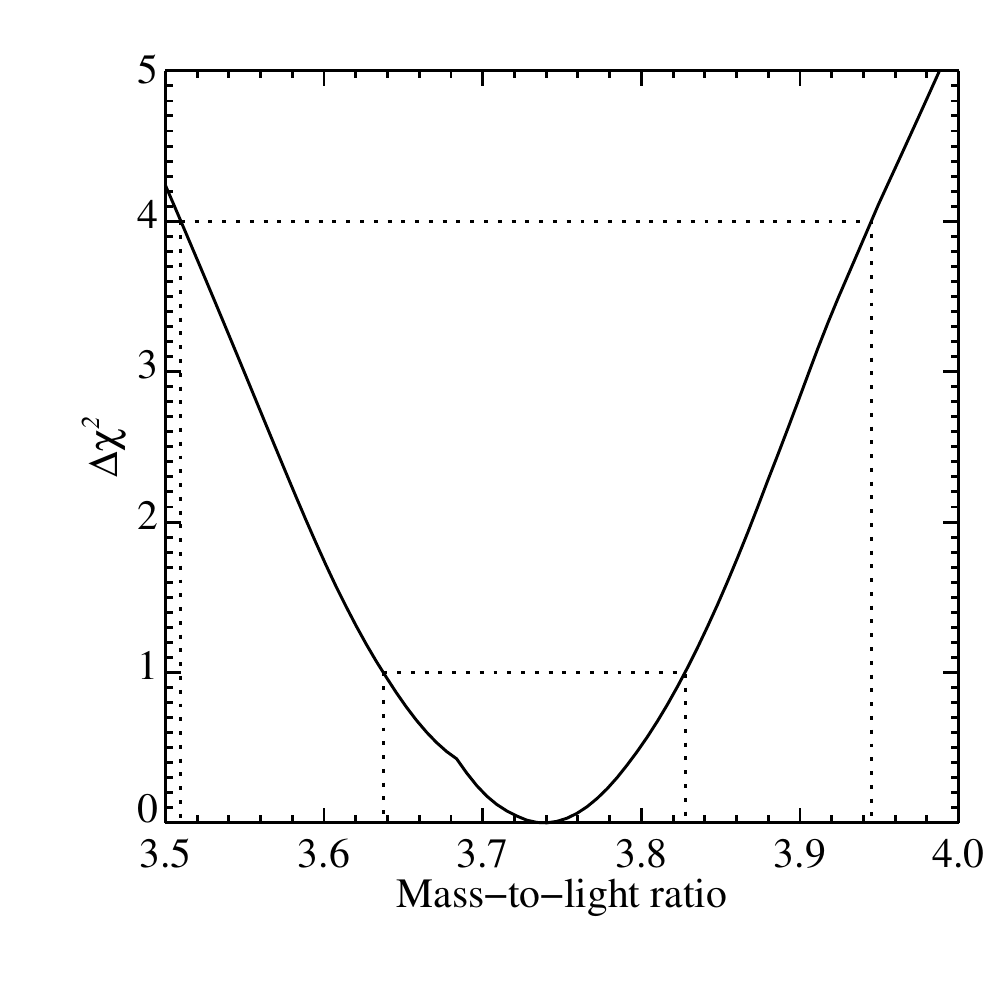}
\caption{Marginalized $\Delta\chi^2$ curves as a function of \mbh\
(\emph{top}) and $\Upsilon_V$ (\emph{bottom}).  Dashed lines show
intersections of the curves with $\Delta\chi^2 =$ 1, 4, and in the
case of the top panel 9, which correspond to the 1, 2, and 3$\sigma$
confidence levels.  There is a slight non-monotonicity in the top
panel as one increases in \mbh\ away from the minimum, but it is small
and entirely within the 1$\sigma$ interval.  Because the deviations
from monotonicity are small (i.e., less than unity), we may be
confident that they are not significant.  In both cases there is a
clear minimum near the best-fit value of each parameter, but in the
case of \mbh, the data are consistent with $\mbh = 0$ at about the
1$\sigma$ level.}
\label{f:margm}
\label{f:margml}
\end{figure}

To determine what values of \mbh\ and $\Upsilon_V$ to consider, we ran a
sparse grid of a wide range of values and then refined around the
best-fitting initial guesses.  The final ranges considered were
$\Upsilon_V = 3.5$--$4.0\ \units{\msun / \lsunv}$ and $\mbh = 0$ to
$2\times10^{8}\units{\msun}$.  The sampling of parameters was not
uniform but may be ascertained from the dots in Figure
\ref{f:n4382chi}.  This figure also plots contours of $\chi^2$ in the
$\mbh$--$\Upsilon_V$ plane.
The best-fit model has parameters $\mbh = 1.5\times10^7\ \msun$ and
$\Upsilon_V=3.72\ \msun / \lsunv$, but models with $\mbh = 0$ are
consistent at about the $1\sigma$ level.  The Gauss-Hermite moments of
the LOSVD for the best-fit model are shown in Figure \ref{f:n4382spec}
and show good agreement with the data.  To obtain our final estimates
for \mbh\ and $\Upsilon_V$, we marginalize over the other parameter to
get $\mbh = 1.3^{+5.2}_{-1.2} \times 10^7\ \msun$ and $\Upsilon_V =
3.74 \pm 0.10\ \msun / \lsunv$, where the errors are 1$\sigma$
uncertainties derived from changes from the minimum marginalized
$\chi^2$ of $\Delta\chi^2 = 1$.  Thus, we cannot rule out the absence
of a black hole, but we may put 2 and 3$\sigma$ upper limits at $\mbh
< 9.6 \times 10^7$ and $\mbh < 1.4 \times 10^8\ \msun$, respectively.
We plot marginalize $\Delta\chi^2$ as a function of \mbh\ and
$\Upsilon_V$ in Figure \ref{f:margm}.

\bookmarksetup{color=[rgb]{0.54,0,0}}
\bookmark[rellevel=1,keeplevel,dest=figchisq]{Fig \ref*{f:n4382chi}: Chi-square contours}
\bookmarksetup{color=black}

\bookmarksetup{color=[rgb]{0.54,0,0}}
\bookmark[rellevel=1,keeplevel,dest=figmargm]{Fig \ref*{f:margm}:
Marginalized delta chi-square.}
\bookmarksetup{color=black}

\begin{figure}
\hypertarget{figaniso}{}%
\centering
\includegraphics[width=\columnwidth]{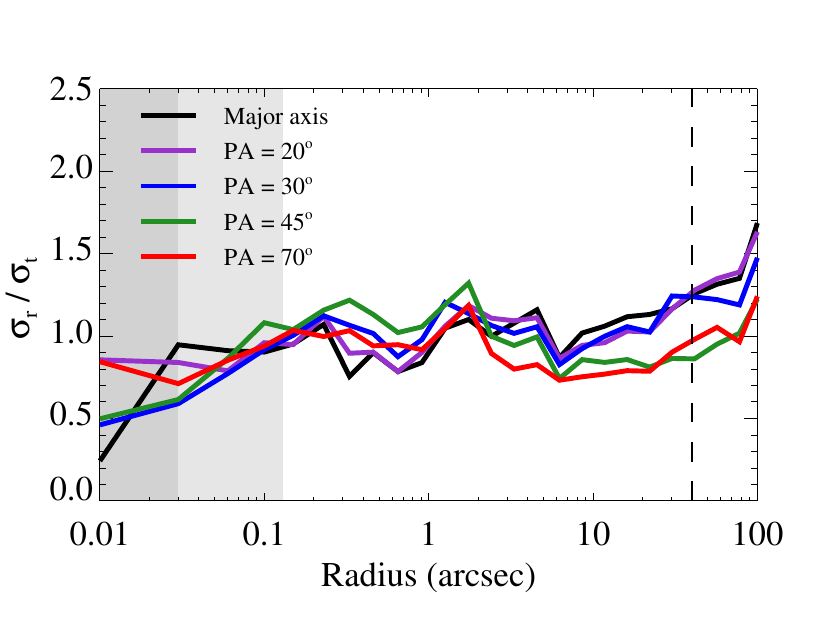}
\caption{Shape of the velocity dispersion tensor for NGC 4382 from the
best-fit model orbit-solution.  The black line is along the major
axis, and the other lines show skew angles with position angles
relative to the major axis as given in the legend.  The values for the
central part of the galaxy are plotted at a radius of 0\farcs01.  The
dashed line shows the radial extent of the ground-based spectroscopic
data.  The orbits are isotropic until the very center at which point
they become tangentially biased.  We indicate the radius of the sphere
of influence of a black hole with mass predicted by the \msigma\
relation (light gray) or with our best-fit mass (medium gray).}
\label{f:n4382aniso}
\end{figure}

\label{anisotropy}
Figure~\ref{f:n4382aniso} shows the velocity dispersion tensor for the
best-fit model by plotting the ratio of the radial velocity dispersion
($\sigma_r$) to the tangential velocity dispersion ($\sigma_t$),
defined as $\sigma_t^2 \equiv (\sigma_\theta^2 + \sigma_\phi^2)/2$,
where $\sigma_\phi$ is the second moment of the azimuthal velocity
relative to the systemic velocity rather than relative to the mean
rotational speed.  An isotropic velocity dispersion would have
$\sigma_r / \sigma_t = 1$.  Uncertainties are $\sim0.2$, estimated
from the smoothness of the profiles \citep{gebhardtetal03}.  There is
a clear transition at 0\farcs1.  Outside of this radius, the orbits
are roughly isotropic ($\sigma_r/\sigma_t \approx 1$), but inside of
this radius the orbits are tangentially biased ($\sigma_r/\sigma_t <
1$).  There is a very strong indication of rotation at the center of
the model galaxy along the major axis ($\sigma_r / \sigma_t = 0.24$).

\bookmarksetup{color=[rgb]{0.54,0,0}}
\bookmark[rellevel=1,keeplevel,dest=figaniso]{Fig \ref*{f:n4382aniso}: Velocity dispersion tensor}
\bookmarksetup{color=black}

\section{Discussion}
\label{concl}

Before discussing the implications of our results, several potential
difficulties in interpretation should be mentioned for this galaxy.
Our kinematic modeling requires three assumptions: (1) a constant
stellar mass-to-light ratio over the range of interest of the galaxy
with the exception of the central black hole, (2) axisymmetric mass
distribution, and (3) that the system is in dynamical equilibrium.  We
discuss each of these in turn.

\subsection{Constant \texorpdfstring{$\Upsilon$}{mass-to-light ratio}?}
\label{caveats}
\label{constantml}
The first of these assumptions, constant mass-to-light ratio, is
unlikely to be far from reality.  Using \emph{XMM-Newton} and
\emph{Chandra} data, \citet{2009A&A...501..157N} study the
gravitational potential as revealed by X-ray emission from the
interstellar medium, assumed to be in hydrostatic equilibrium and
spherically symmetric.  Within $\sim2\units{kpc}$, about
the outer extent of our data, they find a constant $B$-band
mass-to-light ratio consistent with a potential dominated by stellar
mass.  Over the range of $0\farcs2 < r < 10\arcsec$ ($\approx
0.2$--$1\units{kpc}$) NGC 4382 has a shallow $V-I$ color gradient $d(V -
I)/d\log(r) = +0.006 \pm 0.003$, i.e., growing bluer with decreasing
radius, with $V-I = 1.108 \pm 0.001$ at $r = 1\arcsec$
\citep{2005AJ....129.2138L}.  While the sign of the color gradient is
unusual, the magnitude is small enough to dismiss worries about
changing stellar population in the galaxy.

To quantify the effect of deviations from constant mass-to-light ratio
on estimates of black hole mass, we note that when the black hole
kinematic influence is barely resolved, \mbh\ and $\Upsilon$ are
anti-correlated \citep{2009ApJ...700.1690G, 2011ApJ...729...21S}.
Thus if there were unaccounted systematic effects such as a radial
gradient in $\Upsilon$, it would increase the uncertainty in \mbh.  We
can easily estimate the magnitude of the effect since it is linear
with $\Upsilon$.  That is, a 30\% change in $\Upsilon$ will result
at most in a 30\% change in \mbh\ inference.  Based on the radial
color gradient analysis above, the magnitude of any radial $\Upsilon$
gradient must be small, a couple of percent at most.  Thus the total
systematic uncertainty in \mbh\ is less than a couple of percent.

\subsection{Axisymmetric or eccentric disk?}
\label{axisymmetry}
\begin{figure*}
\hypertarget{figimage}{}%
\centering
\includegraphics[height=0.28\textwidth]{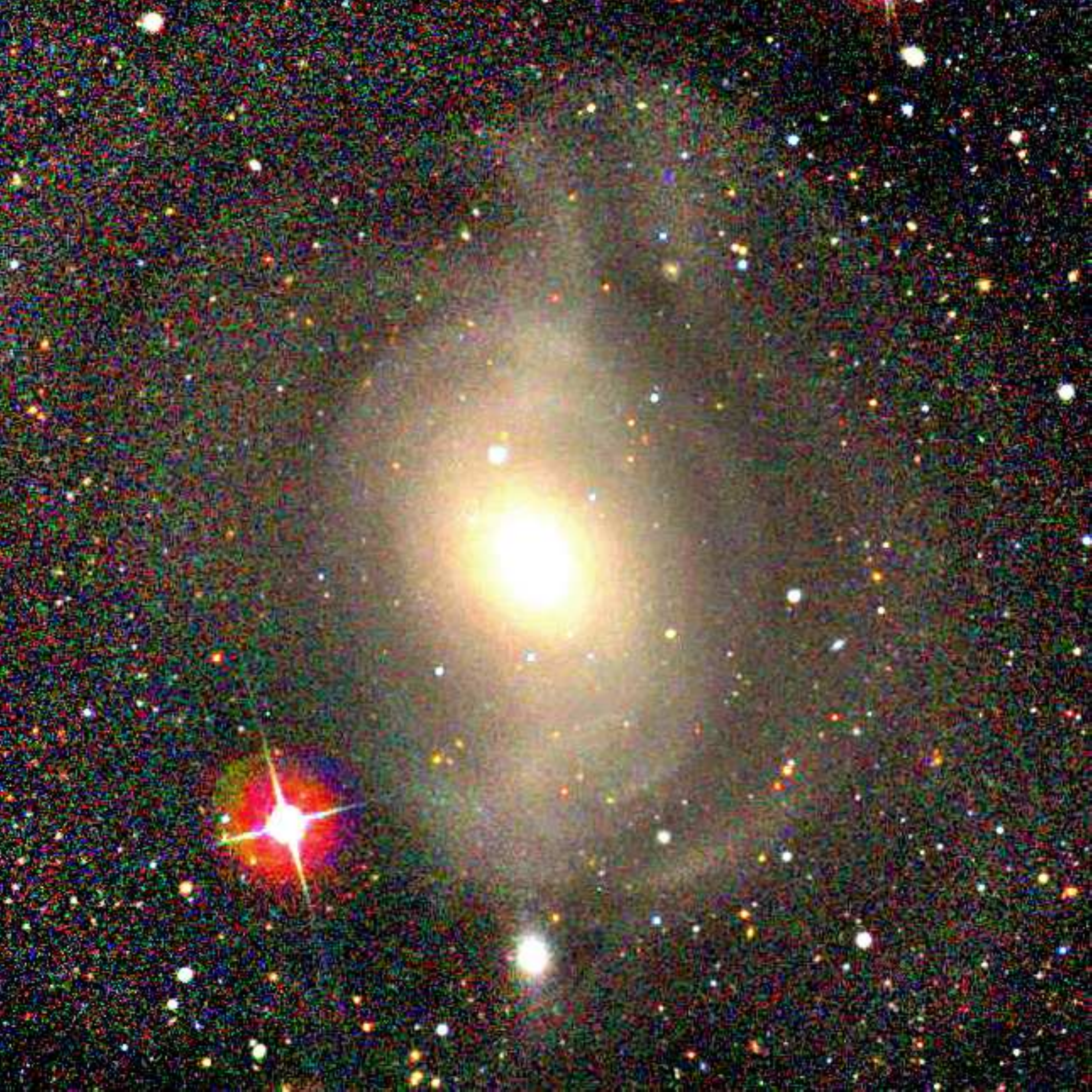}
\includegraphics[height=0.28\textwidth]{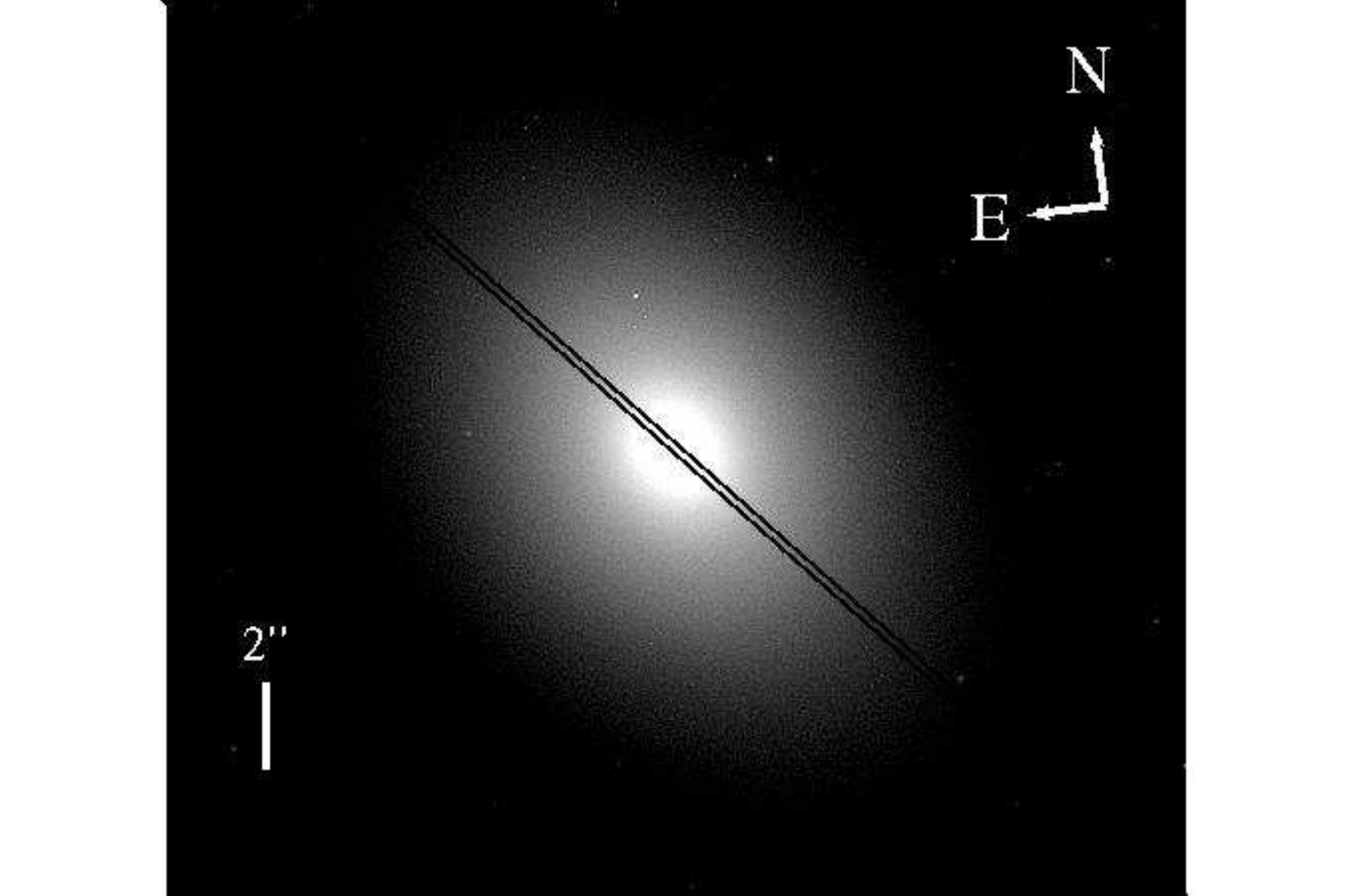}
\includegraphics[height=0.28\textwidth]{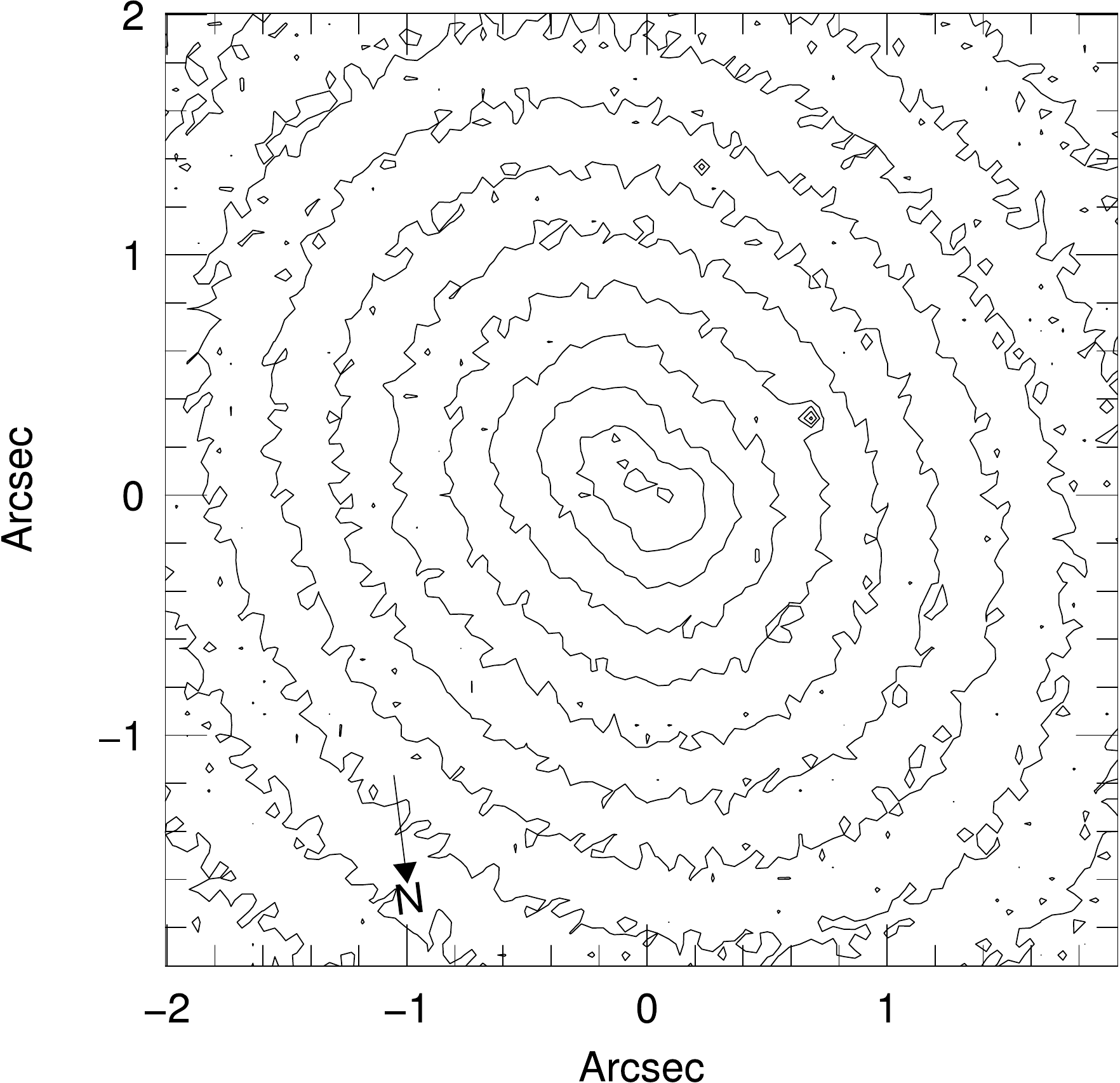}
\caption{Images of NGC 4382.  The left panel shows the
\citet{2009ApJS..182..216K} wide-field image from SDSS $gri$ with a
high-pass filter applied to bring out the fine structure, oriented
such that north is up and east is left.  The center panel is from
F555W \emph{HST}/WFPC image and shows the position of the slit with
regards to the major axis.  The right panel is the
\citet{2005AJ....129.2138L} contour of the same F555W \emph{HST}/WFPC
data zoomed in to show the double-nucleus.  The fine structure and
double nucleus are evidence of a possible recent merger in this
system.  The double nucleus is likely a stellar eccentric nuclear disk
in analogy to M31.}
\label{f:n4382image}
\end{figure*}

The second of these assumptions, axisymmetric mass distribution, is
potentially violated.  Note that the counter-rotating ``kinematically
decoupled core'' (KDC) found in the OASIS data does not necessarily
imply deviations from axisymmetry \citep{2004AN....325..100M}.  As can
be seen from the image of the galaxy (Fig.\ \ref{f:n4382image}), the
change in ellipticity near the center ($\epsilon = 0.6$ at $r \approx
0\farcs2$) to the outer regions ($\epsilon = 0.2$ for $r > 1\arcsec$)
may demonstrate that this system is, overall, triaxial
\citep{2010A&A...510A..45C}.  Modeling a truly triaxial system
requires a triaxial code \citep{2010MNRAS.401.1770V}, but the case for
triaxiality is not straightforward because inside of $0\farcs2$, the
ellipticity decreases again \citep{2005AJ....129.2138L}.

An alternative interpretation of the two-dimensional photometry is
that there are two peaks in surface brightness at the center of the
galaxy with projected separation 0\farcs25
\citep{2005AJ....129.2138L}.  Such a double nucleus appears to be
similar to the double nucleus in M31 (NGC0224;
\citealt{1993AJ....106.1436L}, which \citet{1995AJ....110..628T}
explained as the result of a projection of a central eccentric disk of
stars that is stable only in the presence of a massive dark object,
such as a black hole.

\begin{figure}
\hypertarget{unsymspec}{}%
\centering
\includegraphics[width=\columnwidth]{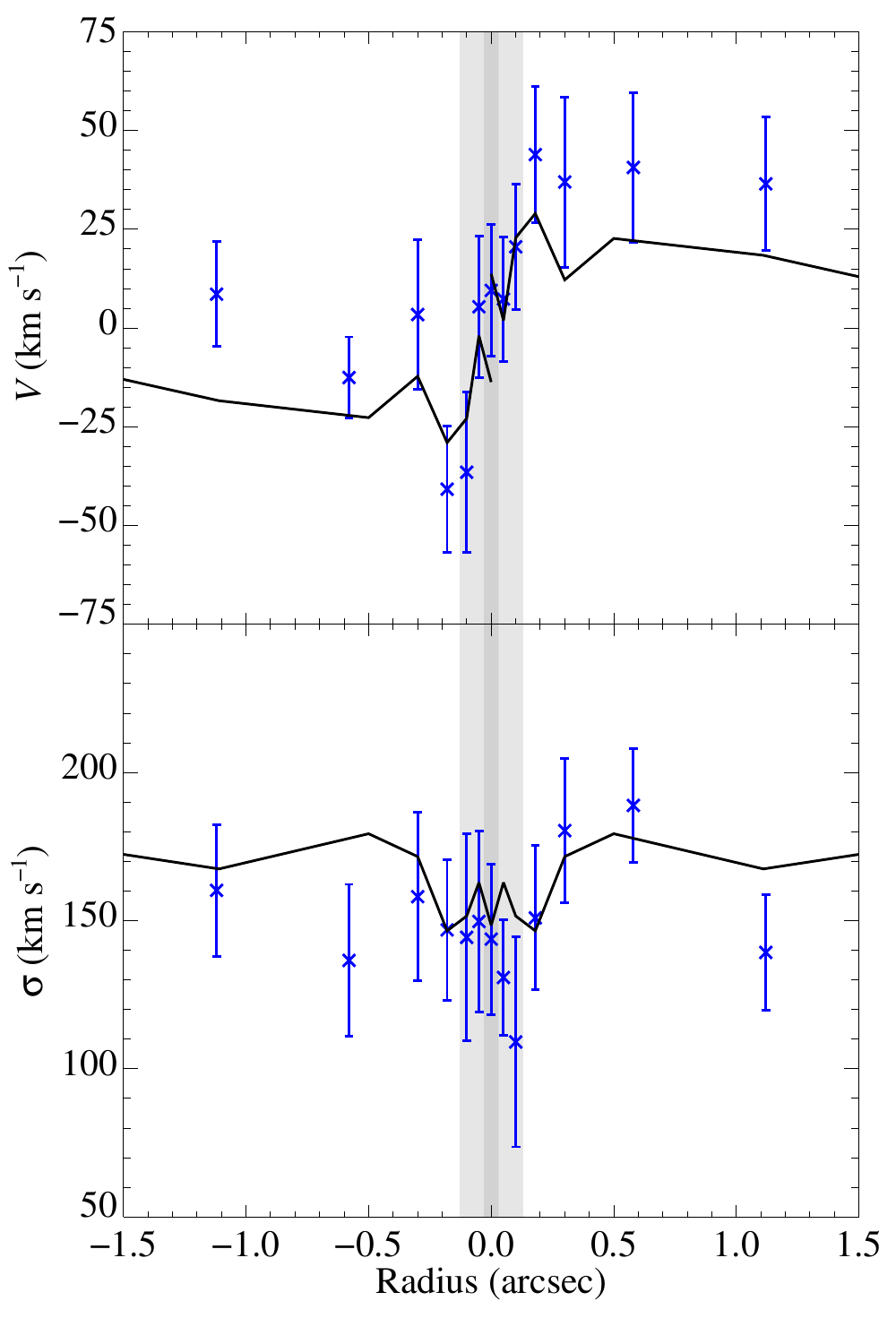}
\caption{Velocity and velocity dispersion measurements from STIS from
both sides of the center.  The black line shows the best-fit
axisymmetric model.  There is a noticeable increase in $\sigma$ at
$R=+0\farcs5$, which corresponds to the location of the secondary
surface brightness peak, indicative of an eccentric stellar disk or
torus.  We indicate the radius of the sphere of influence of a black
hole with mass predicted by the \msigma\ relation (light gray) or with
our best-fit mass (medium gray).}
\label{f:n4382unsymspec}
\end{figure}

This interpretation is born out in the unsymmetrized STIS spectroscopy
(Fig.\ \ref{f:n4382unsymspec}).  There is a prominent increase in the
velocity dispersion at the location of secondary surface brightness
peak $R \approx 0\farcs4$.  This would be expected for an eccentric
stellar disk.

An eccentric disk cannot be modeled by our orbit superposition code,
which forces axisymmetry.  In order to estimate the error introduced
by forcing axisymmetry, we compared the mass that would be obtained by
treating an eccentric disk as though it were a circular orbit.  We use
the mass estimator $\left\langle v_x^2 y\right\rangle$ averaged over
the orbit, where $v_x$ is the line-of-sight velocity and $y$ is the
distance from the center of light of the orbit on the sky.  This is a
suitable surrogate for our modeling.  We then compare the value of the
estimated mass assuming an elliptical orbit to the mass inferred when
assuming a circular orbit with the same semi-major axis.  Treating an
elliptical orbit as though it were a circular orbit produces an error
in the black hole mass that depends on the orbit's eccentricity and
orientation.  For an eccentricity of 0.60, the error in the black hole
mass estimate caused by ignoring the eccentricity varies from $-15\%$
to $+7\%$.  Orbits with smaller eccentricity produce smaller errors.
Given the statistical uncertainties in our \mbh\ estimate, we can
safely ignore this systematic uncertainty.

A consequence of our assumption of axisymmetry is that we are only
sensitive to the presence of a black hole at the center of the galaxy.
There are two potential issues.  The first is whether an existing
black hole is located at the center; the second is whether the STIS
slit was positioned to allow measurement of any central black hole.
We consider both of these.

Because NGC 4382 is a recent merger (see section \ref{virialequil}), a
relevant question is whether any black hole in the galaxy is located
at the center, a requirement for detection with our method.  The
timescale for a black hole to sink to the center of a galaxy depends
on where it starts its decent and in what orbital configuration.  In a
purely tangential orbit at $\sim20$ kpc, it would take longer than a
Hubble time to rest at the center, but a radial orbit at $\sim100$ pc
would take less than $\sim50$ Myr.  In the most likely case, before
the merger, there would be a black hole in the primary galaxy.
Following the merger, it is unlikely that it would have traveled
farther than $\sim100$ pc, and we can expect any such black hole to be
at the center now.

\begin{figure}
\hypertarget{figslit}{}%
\centering
\includegraphics[width=\columnwidth]{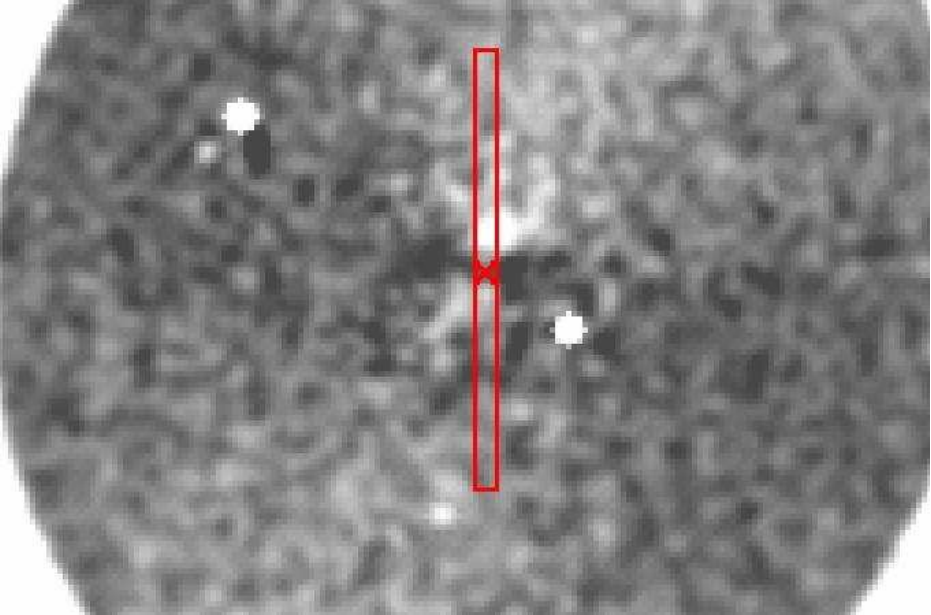}
\caption{Position of STIS slit from the acquisition camera.  This
image comes from the STIS acquisition image taken just before the slit
and dispersive element were added. We have subtracted a symmetric model
for NGC4382 in order to show the non-symmetric features. The red box
shows the location of the slit, and the red cross shows the center
position of the galaxy.  Up is to the southwest and positive $R$ in
Figure \ref{f:n4382unsymspec}.  The figure shows that the slit does
lie on the surface brightness peak.}
\label{f:n4382slit}
\end{figure}

Since the center of NGC 4382 is morphologically complex and we have
not found strong evidence for a black hole at its center, it is
important to determine that the slit was positioned in such a way that
the central kinematics could be determined.  From the STIS acquisition
camera image (Fig.\ \ref{f:n4382slit}), it can be clearly ascertained
that it does so.  This image comes from the STIS acquisition image
taken just before the slit and dispersive element were added. We have
subtracted a symmetric model for NGC4382 in order to show the
non-symmetric features. The red box shows the location of the slit,
and the red cross shows the center position of the galaxy.

\bookmarksetup{color=[rgb]{0.54,0,0}}
\bookmark[rellevel=1,keeplevel,dest=figslit]{Fig \ref*{f:n4382slit}:
Position of STIS slit.}
\bookmarksetup{color=black}

\bookmarksetup{color=[rgb]{0.54,0,0}}
\bookmark[rellevel=1,keeplevel,dest=figimage]{Fig \ref*{f:n4382image}: Images of NGC 4382}
\bookmarksetup{color=black}

\bookmarksetup{color=[rgb]{0.54,0,0}}
\bookmark[rellevel=1,keeplevel,dest=unsymspec]{Fig \ref*{f:n4382unsymspec}: Unsymmetrized LOSVDs.}
\bookmarksetup{color=black}

\subsection{Dynamical equilibrium?}
\label{virialequil}
The third of these assumptions, dynamical equilibrium, is the most
important.  Our analysis, like any dynamical analysis of this kind,
does not handle large deviations from dynamical equilibrium.
\citet{1992AJ....104.1039S} found a high fine structure index in NGC
4382 of $\Sigma = 6.85$, defined as
\beq
\Sigma = S + \log\left(1 + n\right) + J + B + X,
\label{finestructureeq}
\eeq
where $S$ is a visual estimate of the strength of the most prominent
ripples with range $S = 0$--$3$, $n$ is the number of detected
ripples, $J$ is the number of jets, $B$ is a visual estimate of the
maximum boxiness of isophotes, with range $B = 0$--$3$, and $X = 0$ or
$1$ indicates the absence or presence of an $X$-structure,
respectively \citep{1990ApJ...364L..33S}.

\citet{2005AJ....129.2138L} consider NGC 4382 to be an excellent
candidate for a recent merger based on its high value of $\Sigma$, one
the three-highest non-merging in the \citet{1992AJ....104.1039S}
sample, the KDC mentioned above, and the galaxy's blue average $V-I$
color.  \citet{2009ApJS..182..216K} note that the strong
fine-structure features are also seen in the $gri$ SDSS image and
interpret this as evidence that the galaxy has recently merged but has
not fully relaxed.  

The fine structure index, however, is sensitive to gas-rich minor
mergers that are unlikely to disturb the entire galaxy or the central
parts of the galaxy from an equilibrium condition.  Additionally, fine
structure may be a poor measure of recent gas-poor major mergers
\citep{2005AJ....130.2647V, 2009AJ....138.1417T}.  In the case of NGC 4382,
the fine structure appears to be ``Malin shells,'' which can be the
result of gas-poor, primarily stellar, mergers
\citep{1980Natur.285..643M, 1984ApJ...279..596Q}.  As a core
elliptical this galaxy is expected to be a mostly gas-poor merger
\citep{laueretal07b, 2009ApJS..182..216K}.

The virialization time at the center of the galaxy is much shorter
than at the outer regions where the Malin shells are at apopause and
may persist for $\sim10^8\units{yr}$ so that virialization proceeds
from the inside out.  It is thus not likely to affect the inner
portions of the galaxy important for our study
\citep{1990MNRAS.242..311N}.  As a final comment on evidence of recent
merger activity, we note that \citet{2009AJ....138.1741C} investigated
NGC 4382 for signs of interaction motivated by the $\sim35\units{kpc}$
projected distance from NGC 4395.  They found no signatures of
interaction in \ion{H}{1} or other wavelengths.

So we urge caution in over-interpretation of our kinematics modeling,
but for the remainder of this discussion, we take our results at face
value while considering their implications.

\subsection{Resolution of the sphere of influence}

Resolving the sphere of influence of the black hole is desireable but
not necessary for measuring the mass \citep{2009ApJ...698..198G}.  As
the sphere of influence is more and more poorly resolved, the
uncertainties in mass increase until the estimate of black hole mass
is consistent with zero or even negative \citep[e.g., NGC
3945][]{2009ApJ...695.1577G}.  Using our measured effective velocity
dispersion of $\sigma_e = 182\ \kms$, the black hole mass predicted
from the \msigma\ relation is $\mbh = 8.8 \times 10^{7}\ \msun$.  From
equation (\ref{e:rinfdef}) the predicted sphere of influence for this
object at a distance of 17.9 Mpc has radius 0\farcs13 or diameter
0\farcs26.  The central resolution element of our STIS spectroscopy is
0\farcs2 across the slit and 0\farcs05 along the slit.  Thus the
projection of the predicted sphere of influence fits entirely inside
of the central resolution element.  Based on our best-fit mass
estimate of $\mbh = 1.3 \times 10^7\ \msun$, the radius of the sphere
of influence is 0\farcs03 (diameter 0\farcs06) where the the velocity
dispersion is 145 \kms.  The diameter of the actual sphere of
influence is thus about the size of the pixel but 3.3 times smaller
than the slit width.  At such low resolution, it is typical to find an
upper limit to the mass of the black hole, as we have here.  The
1$\sigma$ upper bound to the black hole mass is $\mbh = 6.5 \times
10^7\ \msun$, which has sphere of influence radius 0\farcs13, the same
as the predicted sphere of influence.

\subsection{Is the black hole mass anomalously low?}
Our modeling reveals that the kinematic observations are consistent
with no black hole in NGC 4382 at about the $1\sigma$ level.  If there
were, in fact, no black hole in NGC 4382, it would be one of only two
early-type galaxies with upper limits on their black hole mass below
that predicted by the scaling relations.  The other is NGC 3945, but
the double bars in that galaxy may prevent reliable black hole mass
estimation \citep{2009ApJ...695.1577G}.  Given that our mass
measurement does not completely rule out the presence of a black hole,
we consider whether NGC 4382 is a unique object or merely lies along the
low-end tail of the distribution of black hole masses for a given host
galaxy property.

Based on the \msigma\ relation \citep[$\mbh/\msun = 10^{8.12}
(\sigma_e/200\kms)^{4.24}$;][]{2009ApJ...698..198G}, the mean
logarithmic black hole mass for a galaxy with $\sigma = 182 \kms$ is
$\log{(\mbh/\msun)} = 7.95$, corresponding to $\mbh = 8.8 \times
10^7\ \msun$.  The scatter in the \msigma\ relation is log-normal with
standard deviation $\epsilon_0 = 0.31 \pm 0.06$ for the population of
ellipticals.  Taking the scatter into account, the 68\% confidence
interval of black hole mass is $\mbh = 4. \times 10^7$ to $1.8 \times
10^8\ \msun$, which is consistent within our 1$\sigma$ mass estimate.

To calculate the mass expected from the \ml\ relation
\citep[$\mbh/\msun = 10^{8.95} (L_V /
10^{11}\lsunv)^{1.11}$;][]{2009ApJ...698..198G}, we must first find
the luminosity of the bulge.  The total luminosity of the galaxy may
be obtained from $M_V = -22.54 \pm 0.05$ \citep{2009ApJS..182..216K} using
$\log(L_V/\lsunv) = 0.4 (4.83 - M_V)$
\citep[cf.][]{2008arXiv0807.1393V}.  Given that we are adopting the
more modern classification of E2, this galaxy is ``all bulge,'' and so
the luminosity is $L_V = 8.9 \times 10^{10}\ \lsunv$.  The mean
logarithmic mass for such a galaxy is $\log{(\mbh / \msun)} = 8.89$ or
$\mbh = 7.8 \times 10^8\ \msun$.  Taking the $\epsilon_0 = 0.38 \pm
0.09$ scatter in the relation into account, the 68\% confidence
interval in mass is $\mbh = 3.0 \times 10^{8}$ to $2.1 \times 10^9\
\msun$.  The lower limit of this range is more than a factor of 2
larger than our 3$\sigma$ upper limit for the black hole mass.  In
this sense, the black hole mass is anomalously low.  

In order to be consistent with the \msigma\ reltaion but low according
to the \ml\ relation, NGC 4382 cannot lie on the mean
\citet{1976ApJ...204..668F} $L$--$\sigma$ relation.  In fact, core
galaxies with $M_V = -22.54$ have a mean logarithmic velocity
dispersion corresponding to $\sigma = 253\kms$, compared our measured
value of $\sigma_e = 182\kms$.

The stellar mass of the bulge is $M_* = L_V \Upsilon_V = 3.3 \times
10^{11}\ \msun$.  This is consistent with $g$- and $z$-band model
magnitudes, which give $M_* = 4.0 \times 10^{11}\ \msun$
\citep{2003ApJS..149..289B, 2010ApJ...714...25G}.  Thus $\mbh / M_* =
3.9\times10^{-5}$ compared to the standard $1.3\times10^{-3}$ value
\citep{kg01}.  Using the 3$\sigma$ upper mass bound, the ratio is
$\mbh / M_* = 4.2\times10^{-4}$, still a factor of 3 lower.  In
section \ref{moverl} we discuss how our value of $\Upsilon_V$ may be
wrong because we do not include a dark matter halo in our modeling,
but this is only likely to result in a small (5--10\%) \emph{decrease}
in $\Upsilon_V$ \citep{2009ApJ...700.1690G, 2010arXiv1011.5077S}.

Like many other early type galaxies, NGC 4382 has a flat luminosity
core.  It has been argued that the cores are scoured out by the
inspiral of black holes during galaxy mergers
\citep{1980Natur.287..307B, 1995AstHe..88..238E, 1997AJ....114.1771F,
2003ApJ...582..559V, 2003ApJ...593..661V, 2003ApJ...596..860M,
laueretal07b, 2009ApJ...691L.142K}.  This process ejects stars on
elongated orbits and leads to tangentially biased stellar distribution
functions (See Fig.\ \ref{f:n4382aniso}).  The galaxy's core can be
described by its stellar mass ``deficit,'' the mass in stars ejected
from what was previously a power-law profile.  The most recent
numerical simulations on black hole mergers find that for nearly equal
masses, the mass deficit should scale with the total mass of the
binary \citep{2006ApJ...648..976M}; but for mass ratios far from
unity, the mass deficit should scale with the mass of the secondary
\citep{2008ApJ...686..432S}.  

\citet{2009ApJ...691L.142K} included NGC 4382 in their study of the
correlations between black hole mass and mass deficits.  They
estimated $\mbh = 1.0 \times 10^{8}\ \msun$ by assuming that it
follows the \citet{tremaineetal02} \msigma\ relation, and they
estimated $\Upsilon_V = 8.3$ by assuming that it follows $\Upsilon_V
\propto L_V^{0.36}$ due to \citet{2006MNRAS.366.1126C}.  Using the
photometry from \citet{2009ApJS..182..216K}, they measured a
luminosity deficit of $1.6 \times 10^{8}\ \lsunv$.  For their assumed
values of \mbh\ and $\Upsilon_V$, NGC 4382 lies along the same
correlation between \mbh\ and mass deficit as the rest of their sample
(roughly $M_\mathrm{deficit} = 10 \mbh$).  Using the values for \mbh\
and $\Upsilon_V$ we measure here, NGC 4382 appears to have a small
black hole mass for its core mass deficit, but is consistent at the
1$\sigma$ level.  For our best estimate of the mass,
$M_\mathrm{deficit} / \mbh = 45.6$, and for the 1$\sigma$ upper bound
of the mass $M_\mathrm{deficit} / \mbh = 9.1$.
\citet{2009ApJ...691L.142K} also found that the ratio of light deficit
to total bulge luminosity ($L_{V,\mathrm{deficit}} / L_V =
1.8\times10^{-3}$ for NGC 4382) scales with the ratio of black hole
mass to bulge stellar mass.  Using their empirical correlation and our
measured value of $\mbh / M_*$ predicts a ratio of
$L_{V,\mathrm{deficit}} / L_V = 2.0 \times 10^{-4}$.  With our
1$\sigma$ upper bound on \mbh, it predicts $L_{V,\mathrm{deficit}} /
L_V = 1.7\times10^{-3}$, consistent with the observed value within the
total scatter of the relation.  So it appears that, in sum, the mass
of any black hole in NGC 4382 is consistent with \msigma\ and core
scaling properties but low based on \ml.

The low black hole mass does, however, clear up an apparent disparity
in nuclear activity.  The nuclear X-ray luminosity between 0.3 and 10
keV is less than $L_X < 2.7 \times 10^{38}\ergs$
\citep{2003ApJ...599..218S, 2010ApJ...714...25G}.  The core was also
not detected in radio, typically considered a proxy for jet activity
from an accreting black hole.  With an X-ray limit and a radio
detection, a black hole mass limit can be estimated
\citep{2009ApJ...706..404G}.
Very Large Array observations at 8.4 GHz, however, did not detect any
core radio emission with a $3\sigma$ upper limit of $F_\mathrm{core} <
0.11\units{mJy}$ \citep{2009AJ....138.1990C}.
\citet{2009AJ....138.1990C} note that it is concerning that a galaxy
as large as NGC 4382 shows no sign of nuclear activity.  Their
argument is that, under the assumption that NGC 4382 hosts a $\mbh
\sim 10^8\ \msun$ SMBH as would be expected from galaxy properties,
the absence of nuclear activity in X-ray and radio bands indicates a
failure of AGN activity to trace black holes.  Even a small amount of
ambient gas could produce enough radio emission to be visible for a
large black hole.  Our black hole mass estimate, however, indicates
that it is more likely that NGC 4382 either has no black hole or a
black hole that is far undermassive for its host galaxy properties.
Therefore, the lack of any nuclear activity accurately traces the
small or non-existent black hole in line with predictions for and
existing observations of the AGN fraction in Virgo cluster galaxies
\citep{2008MNRAS.384.1387V, 2008ApJ...680..154G}.

\subsection{Mass-to-light ratio}
\label{moverl}
Although this work does not unambiguously detect a black hole, the
mass-to-light ratio was determined precisely, $\Upsilon_V = 3.74 \pm
0.10\ \msun / \lsunv$ owing to the good coverage and high quality of
the SAURON data.  Unfortunately, NGC 4382 was not one of the galaxies
modeled by \citet{2006MNRAS.366.1126C}; so no direct comparison can be
made.  Note that we do not explicitly include a dark matter halo in
our modeling.  This is unlikely to affect our black hole mass results
on account of the high spatial resolution provided by \emph{HST}/STIS
\citep{gebhardtetalm87}.  The stellar mass-to-light ratio, however, is
likely to be affected by not including this component.  While the
inferred stellar mass-to-light ratio changed by a factor of 2 when
including or omitting a dark matter component in the modeling for M87
\citep{2009ApJ...700.1690G}, the high spatial resolution in our data
sets is likely to result in only a small (5--10\%) decrease in
$\Upsilon_V$ \citep{2010arXiv1011.5077S}.  Galaxies with $M_V =
-22.54$ have a mean mass-to-light ratio $\Upsilon_V = 6.7$
\citep{laueretal07}.  Based on the scaling relation between
$\Upsilon_I$ and stellar velocity dispersion in ellipticals
\citep{2006MNRAS.366.1126C}, the implied typical value for an galaxy
with $\sigma_e = 182\kms$ is roughly $\Upsilon_V \approx 5.3$.  The
low value of $\Upsilon_V$ in NGC 4382 compared to established
ellipticals suggests ongoing star formation and would be expected for
a recent merger.

\hypertarget{ackbkmk}{}
\acknowledgements We thank Scott Tremaine for helpful discussions
about this work and the SAURON team for making their data available to
the community.  We thank the anonymous referee for his or her useful
comments.  The modeling made use of the facilities at the Texas
Advanced Computing Center at the University of Texas at Austin.  This
work made use of the NASA's Astrophysics Data System (ADS), and the
NASA/IPAC Extragalactic Database (NED), which is operated by the Jet
Propulsion Laboratory, California Institute of Technology, under
contract with the National Aeronautics and Space Administration. This
research has made use of the VizieR catalogue access tool, CDS,
Strasbourg, France.  Financial support was provided by NASA/\emph{HST}
grants GO-7468, and GO-9107 from the Space Telescope Science
Institute, which is operated by AURA, Inc., under NASA contract NAS
5-26555.  
KG\"u acknowledges support provided by the National Aeronautics and Space
Administration through Chandra Award Number GO0-11151X issued by the
Chandra X-ray Observatory Center, which is operated by the Smithsonian
Astrophysical Observatory for and on behalf of the National
Aeronautics Space Administration under contract NAS8-03060.
DOR thanks the Institute for Advanced Study and acknowledges
support of a Corning Glass Works Foundation Fellowship.  
KGe acknowledges support from NSF-0908639.
\bookmark[level=0,dest=ackbkmk]{Acknowledgements}

\bibliographystyle{apjads}
\hypertarget{refbkmk}{}%
\bookmark[level=0,dest=refbkmk]{References}
\bibliography{gultekin}

\label{lastpage}
\end{document}